\documentclass[sigconf]{acmart}

\usepackage{booktabs} % For formal tables
\usepackage{transparent}

\usepackage{graphicx}
\usepackage{caption}
\usepackage{subcaption}
\usepackage{ftnxtra}

\setcopyright{rightsretained}
\begin{document}
\title{Recommending Complementary Products in E-Commerce Push Notifications with a Mixture Model Approach}

\author{Huasha Zhao, Luo Si, Xiaogang Li, Qiong Zhang} 
    \affiliation{ 
      \institution{Alibaba Group}
      \streetaddress{400 S El Camino Real, \#400}
      \city{San Mateo} 
      \state{California, USA} 
      \postcode{94402}
    }
    \email{  {huasha.zhao, luo.si, xiaogang, qz.zhang }@alibaba-inc.com}

\begin{abstract}
Push notification is a key component for E-commerce mobile applications, which has been extensively used for user growth and engagement. The effectiveness of the push notification is generally measured by message open rate. A push message can contain a recommended product, a shopping news and etc., but often only one or two items can be shown in the push message due to the limit of display space. This paper proposes a mixture model approach for predicting push message open rate for a post-purchase complementary product recommendation task. The mixture model is trained to learn latent prediction contexts, which are determined by user and item profiles, and then make open rate predictions accordingly. The item with the highest predicted open rate is then chosen to be included in the push notification message for each user. The parameters of the mixture model are optimized using an EM algorithm. A set of experiments are conducted to evaluate the proposed method live with a popular E-Commerce mobile app. The results show that the proposed method is superior than several existing solutions by a significant margin.
\end{abstract}

%\begin{CCSXML}
%<ccs2012>
%<concept>
%<concept_id>10002951.10003317</concept_id>
%<concept_desc>Information systems~Information retrieval</concept_desc>
%<concept_significance>500</concept_significance>
%</concept>
%<concept>
%<concept_id>10010405.10003550</concept_id>
%<concept_desc>Applied computing~Electronic commerce</concept_desc>
%<concept_significance>500</concept_significance>
%</concept>
%</ccs2012>
%\end{CCSXML}

%\ccsdesc[500]{Information systems~Information retrieval}
%\ccsdesc[500]{Applied computing~Electronic commerce}
%\keywords{Push Notification, E-Commerce, Mixture Model, Probabilistic Latent Class Model}
\copyrightyear{2017} 
\acmYear{2017} 
\setcopyright{acmcopyright}
%\acmConference{SIGIR '17}{August 07-11, 2017}{Shinjuku, Tokyo, Japan}
\acmConference{SIGIR'17}{}{August 7--11, 2017, Shinjuku, Tokyo, Japan.}
\acmPrice{15.00}\acmDOI{http://dx.doi.org/10.1145/3077136.3080676}
\acmISBN{978-1-4503-5022-8/17/08}

\maketitle
\vspace{-0.15cm}
\section{Introduction}

%    \begin{figure}
%    \centering
%    \subfigure [Transaction related push. The above message reminds the user that the purchased teapot has been shipped from seller.]{\includegraphics[width=0.10\textwidth]{push1_meitu.png}}
%
%    \subfigure [B]{\includegraphics[width=0.10\textwidth]{push2_meitu.png}}
%
%    \caption{Two examples of push notification messages.}
%    \label{pm}
%    \end{figure}

%\begin{figure}
%    \centering
%    \begin{subfigure}[b]{0.25\textwidth}
%        \includegraphics[width=\textwidth]{push1_meitu.png}
%        \caption{Transaction related push message. The above message reminds the user that the purchased XianMingJu health teapot has been shipped from seller.}
%    \end{subfigure}
%    \begin{subfigure}[b]{0.25\textwidth}
%        \includegraphics[width=\textwidth]{push2_meitu.png}
%        \caption{Marketing push message. The above message recommends TieGuanYin (a type of tea leaf) after the user purchased a teapot.}
%    \end{subfigure}
%    \caption{Two examples of push messages}
%    \label{pm}
%\end{figure}

\begin{figure}
\centering
\includegraphics[width=0.19\textwidth]{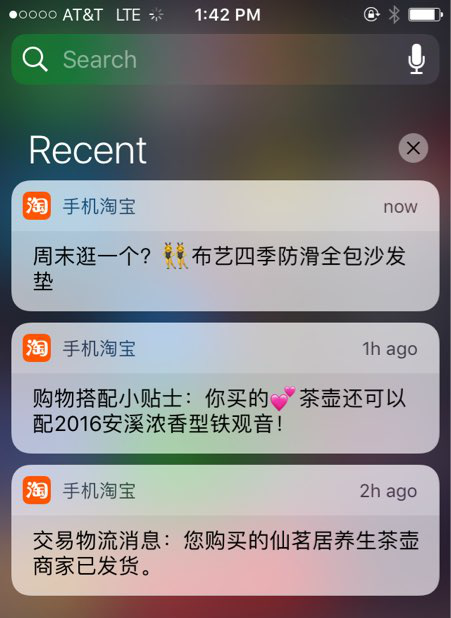}
\caption{Three examples of push messages {\footnotemark}: The top two are marketing push messages. The top message is a PPR which recommends a personalized sofa mat without making specific connections with user behavior. The middle message is a CPR which recommends Tie Guan Yin tea after the user purchased a teapot. The bottom one is transaction related push message. The message reminds the user that the purchased teapot has been shipped from seller. }
\label{pm}
\vspace{-0.4cm}
\end{figure}

Push notification service \cite{bell2011push,tan2016exploration, kumar2015push, moertini2012commerce} is a key component of E-commerce mobile applications. It pushes text messages to users to provide a seamless shopping experience. The push messages are usually classified into two categories: transaction related push message which reminds the user of transaction information on existing orders, such as payment processing, logistic updates and etc., and marketing push message which promotes new product or shopping news that the user may be interested in. Some examples of push messages are illustrated in Figure \ref{pm}. Push message has been extensively used for user growth and user engagement. It initializes the connection with user in an active manner, and can bring users back to the E-commerce mobile app. The effectiveness of the push message is measured by {\em open rate} which is the percentage of messages clicked/viewed by users.

In marketing push notification, recommending the ``right'' product is critical. There has been substantial research in recommendation systems \cite{herlocker1999algorithmic, su2009survey, bobadilla2013recommender}. However, push notification has several key differences in comparison with classic recommendation tasks. First, the text template or the presentation of the push message can be more important in determining the open rate than the recommended product itself. Attractive message slogans sometimes play a decisive role. To increase open rate, the message needs to be created with strong and preferably direct attachment with the user behavior. Furthermore, only one or two items can be shown in the push message due to the limit of display space. Finally, only a given number of messages can be sent to a user during a period of time, to avoid spamming. As a result, the prediction accuracy of open rate becomes critically important due to limited chance of recommendation.\footnotetext{Translations of the messages in the figure, from top to bottom. 1, Go shopping during the weekend? All-season all-purpose non-slip cloth sofa mat for you. 2, Shopping together: 2016 Anxi fragrant Tie Guan Yin tea with your newly bought teapot. 3, Logistic reminder: your purchased item Xianmingju health teapot has been shipped. }

In this paper, we propose a method for recommending post-purchase complementary products through push notification. Post-purchase recommendation has significant advantage in the push setting in comparison with pushing a single personalized product without connecting it to user history. Examples of complementary product recommendation (CPR) and personalized product recommendation (PPR) are illustrated in Figure \ref{pm}. PPR pushes only one product to the user using some collaborative filtering algorithms, and the message contains only one anchor to the user. In contrast, CPR naturally embeds two anchors in the push message template - the purchased product and the recommended product, and it offers stronger connections with the user than PPR. As a result, such messages are more attractive to users. 

In addition, a mixture model is presented to predict message open rate. The predicted open rate is in turn used to determine the product to recommend. Mixture model is also known as probabilistic latent class model, and it has been used in user modeling and recommendation systems \cite{jin2006study, cetintas2013probabilistic, cetintas2013forecasting, tao2004two}. A mixture model can automatically learn the underlying structure of a prediction task, and improve the prediction accuracy by training separate sets of forecasting weights for different latent prediction classes.

In our problem, the mixture model is used to represent hidden prediction contexts. A context is determined by a combination of user and product profiles. For example, some users may be more interested in the popularity of the product, while others care more about the complementarity between the recommended product and purchased product. In this case, in the prediction model we need to assign higher weights on product popularity feature to the first group of users while give more weights on complementary score to the second group. A mixture model can be trained to learn such latent contexts for prediction.

The rest of the paper is arranged as follows: Section 2 describes complementary score calculation for CPR, and the score is one of the key features used in the model. The mixture model for open rate prediction is discussed in Section 3. Section 4 presents experiment results along with analysis. And finally Section 5 concludes the paper.

\vspace{-0.3cm}
\section{Complementary Product Recommendation}
\label{cpr}

Complements and substitutes are two types of recommendation products \cite{mcauley2015inferring, lu2015competition}. Complements can be bought in addition to each other. However, substitutes can be bought instead of each other. In the post-purchase recommendation scenario, it is critical to recommend complementary products, but not substitutes. This reason is that users have already made the decision, and may feel being spammed if a product that serves the same purpose is recommended again.  Here we present the method we use for complementary product candidate selection. We choose the product pairs with high co-purchase scores and low substitutivity scores at the same time. The scoring method is described in the following.
\vspace{-0.3cm}
\subsection{Co-Purchasing Graph}
Assume $A_{ui}$ is the user-product purchase graph for user $u \in {\mathbb U}$ and product $i \in {\mathbb P}$ over a certain period of time, and $t(A_{ui})$ is the timestamp at which the purchase event happens. The co-purchase score $p_{ij}$ - the score for buying $j$ after buying $i$ is defined as follows,
\begin{align}
p_{ij} = \frac{\sum_{u \in {\mathbb U}} A_{ui} A_{uj} {\mathbf 1}(t(A_{uj}) > t(A_{ui}))}{\sqrt{\sum_{u \in {\mathbb U}} A^2_{ui} \sum_{u \in {\mathbb U}} A^2_{uj} }}.
\end{align}

\subsection{View-and-then-Purchase Graph}
Further assume $B_{ui}$ is the user-product view graph, and $t(B_{ui})$ is the timestamp at which the view event happens. The substitutivity score $q_{ij}$ between product $i$ and  $j$ is defined as follows,
\begin{align}
q_{ij} = \frac{\sum_{u \in {\mathbb U}} B_{ui} A_{uj} {\mathbf 1}(t(A_{uj}) > t(B_{ui}))}{\sqrt{\sum_{u \in {\mathbb U}} B^2_{ui} \sum_{u \in {\mathbb U}} A^2_{uj} }}.
\end{align}

The final complementary score for product pair $i, j$ is determined by,
\begin{align}
s_{ij} = p_{ij} - q_{ij}.
\label{cs}
\end{align}
The score selects product pairs with high complementarity and low substitutivity. We also calculate complementary scores for categories with the same method.

\section{Mixture Model for Open Rate Prediction}
In this section, the mixture model for predicting the message open rate for user-product pairs is presented. The motivation comes from the nature of users and products. For instance, some users prefer popular products (product popularity score) regardless of the complementariness between the two products  (complementary score). At the same time, different classes of products may also cause different sets of prediction parameters.

At high level, the model is comprised of two parts: an assignment model which maps the inputs to prediction {\it contexts} and a context-aware prediction model for open rate forecasting. Formally, we assume the following probabilistic model:
\begin{align}
\nonumber
P({\mathbf y} | X, {\hat X}, \Theta, \Psi) &= \prod_{i=1}^{N} P(y_i | {\mathbf x_i},{\mathbf {\hat x}}_i, \Theta, \Psi)  \\
&= \prod_{i=1}^{N} \sum_{z_i=1}^{M}P(z_i | {\mathbf {\hat x}}_i, \Theta) P(y_i | {\mathbf x_i}, z_i, \Psi). 
\end{align}
Here $N$ is the total number of examples in the dataset. There are two sets of features ${\mathbf {\hat x}}_i \in \mathbb{R}^{m}, {\mathbf x_i} \in \mathbb{R}^{n}$ for each example $i \in \{1, \dots, N\}$. ${\mathbf {\hat x}}_i$ is the context assignment feature and ${\mathbf x_i}$ is the open rate prediction feature. In practice, features in these two sets may have overlaps. We describe these features in more details in Section \ref{feat}. Furthermore, each example is labeled $y_i \in \{0, 1\}$ to represent whether it is opened or not,  and a hidden variable $z_i$ which assigns each example to a predicting context.  We further assume $M$ contexts of interest and  $z_i \in \{1, \dots, M\}$. The model is characterized by two sets of parameters, i.e. $\Theta = (\theta_1, \dots, \theta_M)$ and  $\Psi = (\psi_1, \dots, \psi_M)$. Both $\theta_k$ and $\psi_k$, $k \in \{1, \dots, M\}$ are of the same dimension as ${\mathbf {\hat x}}_i$ and ${\mathbf x_i}$ respectively, for $ i \in \{1, \dots, N\}$, and they parameterize the assignment model and prediction model of the $k^{th}$ context respectively.

In more details, the assignment process can be modeled with a multi-class logistic model as the following, 
\begin{align}
\nonumber
P(z=k | {\hat {\mathbf x}} = {\hat x} ) = & \frac {\exp(\theta_k^t {\hat x})} {1 + \sum_{j=1}^{M-1}\exp(\theta_j^t {\hat x})},  \quad k = 1,  \dots, M-1, \\
P(z=M & | {\hat {\mathbf x}} = {\hat x} ) = \frac {1} {1 +  \sum_{j=1}^{M-1}\exp(\theta_j^t {\hat x})}. 
\end{align}
In this case, $\Theta$ is only defined up to a multiplicative constant, so that $\theta_M$ can be omitted in the model. A binary logistic regression model
is used for prediction, for each scenario $k \in \{1, \dots, M\}$ respectively,
\begin{align}
\nonumber
P(y=0 | {\mathbf x} = x, z=k) = \frac {\exp(\psi_k^t x)} {1 +\exp(\psi_k^t x)},  \\
P(y=1 | {\mathbf x} = x, z=k) = \frac {1} {1 + \exp(\psi_k^t x)}.  
\end{align}
The log-likelihood expression for model from the data $(X, {\mathbf y})$ is difficult to optimize because it involves the log of the sum.  If the hidden variable $z$ is introduced, however, the likelihood function can be significantly simplified:
\begin{align}
\log({\mathcal L}(\Theta, \Psi | X, {\hat X},  {\mathbf y}, {\mathbf z})) &= \log (P(X, {\hat X}, {\mathbf y}, {\mathbf z} | \Theta, \Psi )) \\\nonumber
&= \sum_{i=1}^{N} \log(P({\mathbf x_i})P({\mathbf {\hat x}}_i)P(z_i | {{\mathbf {\hat x}}_i}, \Theta) P(y_i | {\mathbf x_i}, z_i, \Psi)).
\end{align}
The above likelihood is intractable for solving optimized parameters analytically. We resort to the EM algorithm to find optimal $\Theta$ and $\Psi$. In the E step, we first derive an expression of the posterior distribution of the unobserved data. Using Bayes's rule and property of conditional independence, we can compute, 
\begin{align}
P(z_i|{{\mathbf {\hat x}}_i, \mathbf x_i}, y_i, \Theta, \Psi) = \frac {P(z_i | {\mathbf {\hat x}}_i, \Theta) P(y_i | {\mathbf x_i}, z_i, \Psi)}{\sum_{k=1}^{M}P(k | {\mathbf {\hat x}}_i, \Theta) P(y_i | {\mathbf x_i}, k, \Psi)}.
\end{align}
An auxiliary Q function \cite{bilmes1998gentle} of the likelihood can be derived as,
\begin{align}
\nonumber
Q&(\Theta^\prime, \Psi^\prime | \Theta, \Psi) \\\nonumber
=& E[\log({\mathcal L}(\Theta^\prime, \Psi^\prime | X, {\hat X},  {\mathbf y}, {\mathbf z})) | X, {\hat X}, {\mathbf y}, \Theta, \Psi] \\\nonumber
%=& \sum_{{\mathbf z} \in {\mathcal Z}}  \sum_{i=1}^{N} \log(P(z_i | {\mathbf {\hat x}}_i, \Theta^\prime) P(y_i | {\mathbf x_i}, z_i, \Psi^\prime)) \prod_{i=1}^{N}P(z_i|{\mathbf {\hat x}}_i, {\mathbf x_i}, y_i, \Theta, \Psi)  \\\nonumber
=& \sum_{k=1}^{M} \sum_{i=1}^{N} \log(P(k | {\mathbf {\hat x}}_i, \Theta^\prime) ) P(k | {\mathbf {\hat x}}_i, {\mathbf x_i}, y_i, \Theta, \Psi) \\
+& \sum_{k=1}^{M} \sum_{i=1}^{N} \log(P(y_i | {\mathbf x_i}, k, \Psi^\prime) ) P(k | {\mathbf {\hat x}}_i, {\mathbf x_i}, y_i, \Theta, \Psi). 
\end{align}
Note, $P({\mathbf x_i})$ and $P( {\mathbf {\hat x}}_i)$ are removed in the Q function since we assume they both follow uniform distribution. To maximize the the Q function, we can maximize the term containing $\Theta^\prime$ and the term containing $\Psi^{\prime}$ independently since they are not related. We can also solve  $\theta_k^\prime$ for each scenario independently for the same argument. The M step updates can be derived as the following,
\begin{align}
\Theta^\prime \propto \operatorname{arg\,max}_{\Theta^\prime}   \sum_{k=1}^{M} \sum_{i=1}^{N} \log(P(k | {\mathbf {\hat x}}_i, \Theta^\prime) ) P(k | {\mathbf {\hat x}}_i, {\mathbf x_i}, y_i, \Theta, \Psi) , \label{m1} \\
\psi_k^\prime \propto \operatorname{arg\,max}_{\psi_k^\prime} \sum_{i=1}^{N} \log(P(y_i | {\mathbf x_i}, k, \psi_k^\prime) ) P(k | {\mathbf {\hat x}}_i, {\mathbf x_i}, y_i, \Theta, \psi_k). \label{m2}
\end{align}
The above updates can be solved using a gradient decent solver.

\thispagestyle{empty}

\subsection{Model Features}
\label{feat}
As shown in Table \ref{mf}, there are 4 types of features we use in the mixture model. They are user features, product features, user-product features and product-product features. 

Besides user demographics, we also use user cluster features and user active scores. User clusters are generated by running k-means on user shopping behaviors (at category level) within certain time period. User active score measures how active the user is on the E-Commerce platform. Furthermore, we use product features and user-product features generated from multiply time intervals, to capture the time dynamic of user and product behaviors. Finally, we also include product-product features: the calculations of which are presented in Section \ref{cpr}.

User features and product features are included in ${\mathbf {\hat x}}_i$, to predict prediction context. While all features except user features are used in the second stage (open rate prediction) of the mixture model (in ${\mathbf {x}}_i$) - user features are not needed when ranking products for a specific user.

\begin{table}
\centering
\begin{tabular}{ | p{2.8cm} | p{5cm} | }
\hline
user features & user cluster id \\
\cline{2-2}
 & user active score \\
\cline{2-2}
 & user demographics, e.g. age, income and etc. \\
\hline
product features &  product sales  in the past 1, 2, 7, 28 days\\
\cline{2-2}
 & product views  in the past 1, 2, 7, 28 days\\
\cline{2-2}
 & price and other metadata \\
\hline
user-product features & user-product preference scores in the past 1, 2, 7, 28 days respectively \\
\cline{2-2}
& user-category preference scores in the past 1, 2, 7, 28 days respectively \\
\hline
product-product features & product complementarity scores defined in equation \ref{cs} \\
\cline{2-2}
& category complementarity scores \\
\hline
\end{tabular}
\caption{Model features.}
\label{mf}
\vspace{-0.8cm}
\end{table}

\section{Experiments}
This section presents the experiments for evaluating the mixture model. The model is tested live on a real-world push notification task on a popular E-commerce mobile application with hundreds of millions of active users.

\subsection{Experiment Setup}

In model training, one month push notification log data is used. The complementary product push task reaches 10 million users daily.  There are in total 300 million records for training. L-BFGS solver is applied to compute equations (\ref{m1}) (\ref{m2}) required by the M-step. The threshold of likelihood convergence is set to $1e^{-5}$.

\subsection{Number of Contexts Evaluation}

\begin{figure}
  \centering
   \includegraphics[width=0.335\textwidth]{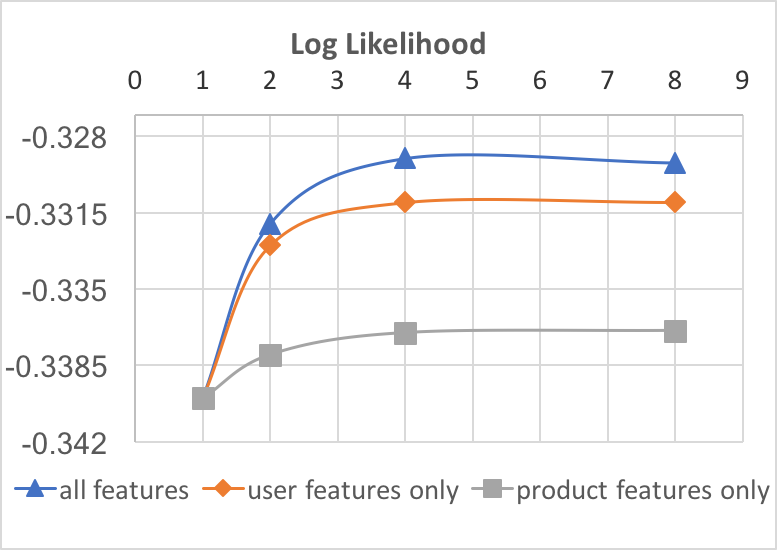}
   \caption{Log-likelihood per example with different feature sets and different number of contexts $k$.}
   \label{ll}
   \vspace{-0.2cm}
\end{figure}

The optimal number of hidden contexts $k$ is evaluated in this section. Figure \ref{ll} plots the log likelihood at convergence for different $k$. As we can see, there is no benefit of increasing $k$ beyond $k=4$. In addition, we also evaluate the model performance for different sets of features (${\mathbf {\hat x}}_i$) used for context prediction. We compare full features (described in Section \ref{feat}) with user features only and product features only. The full model learns the user and product mixture jointly. As a result, it outperforms both user only and product only models. From Figure \ref{ll}, we also observe that the model of product features only is much worse. This means there are more opportunities to explore user mixtures than product mixtures. In other words, it will be more beneficial to model open rate at user dimension rather than at product dimension, if we have to choose one. 

\subsection{Impact of User Cluster on Model Weights}
\begin{figure}
  \centering
   \includegraphics[width=0.325\textwidth]{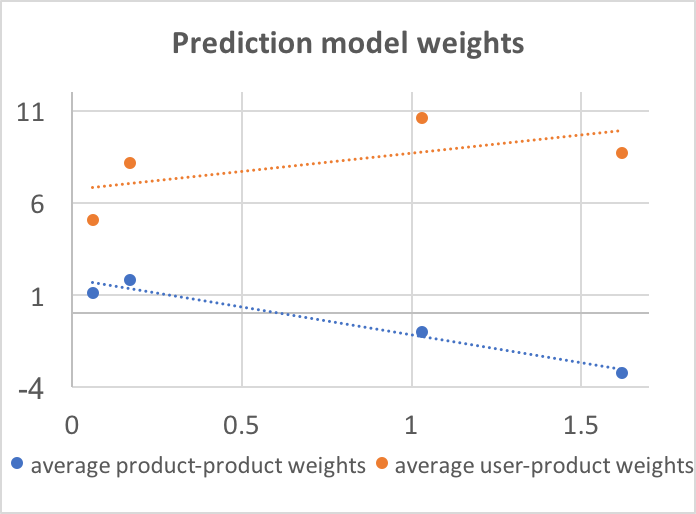}
   \caption{Prediction model weights for different user clusters based on user active scores.}
   \label{ana}
   \vspace{-0.2cm}
\end{figure}

Experiments are run to illustrate the impact on prediction model weights of different sub-populations in the mixture model. Figure \ref{ana} plots the average of product-product feature weights and user-product feature weights in the prediction model against the weight of user active score in the context assignment model. Assignment model with higher user active score weight selects more active users. As shown in the figure, more active users prefer products with higher user preference scores, while less active users prefer more on complementariness between products. Intuitively this can be explained as: sophisticated users know what they want to buy and have strong product preferences, while newbie users rely more on the complementary product recommendation from the platform.

\subsection{Online Experiment Results}
The CPR with mixture model is deployed live to our production push recommendation system, and A/B tests are conducted. Results are shown in Table \ref{or}. The baseline just pushes the most popular products to users without any personalization. In PPR, a state-of-the-art item-based recommender is applied to compute personalized product. The baseline CPR model selects the product for a user with highest user-product score times product complementarity score without considering other features. It is also worth noting that a mixture model (MM) with $k=1$ is equivalent to a logistic regression model for open rate prediction.

CPR outperforms PPR by a significant margin. This confirms our hypothesis discussed in the introduction. The CPR textual presentations are richer and contain two anchors to the users. And therefore, CPR delivers better results. The mixture model offers additional gains in message open rate. The open rate improvement is around 11\%. The figure also shows it is beneficial to model latent user product clusters for open rate prediction. In comparison with a plain logistic regression (mixture model with $k=1$), the gain is about 3\%. Although 3\% does not look like a very big number, the impact of the mixture model is significant given the huge user base.

%\begin{figure}
%  \centering
%   \includegraphics[width=0.4\textwidth]{openrate.png}
%   \caption{Open rate relative to baseline}
%   \label{or}
%\end{figure}

%\begin{table} [H]
%\centering
%\begin{tabular}{ c c c c c }
%\hline
%\hline
%              & PPR & CPR & \begin{tabular}{@{}c@{}}CPR with\\ MM k=1\end{tabular} &  \begin{tabular}{@{}c@{}}CPR with \\ MM k=4\end{tabular}\\
%\hline
%\begin{tabular}{@{}c@{}}Open rate rel. \\ to baseline\end{tabular}   & 2.01 & 2.89 & 3.10 & 3.22   \\
%\hline
%\begin{tabular}{@{}c@{}}P-value against \\  prev. model\end{tabular}  &  $.00011^{***}$ &  $.00036^{***}$ & $.00073^{***}$ & $.0046^{**}$ \\
%\hline
%\end{tabular}
%\caption{Open rates. P-values are calculated to measure the statistical significancy against the {\em previous} model.}
%\label{or}
%\vspace{-0.5cm}
%\end{table}

\begin{table}
\centering
\begin{tabular}{ |c |c |c| }
\hline
              & \begin{tabular}{@{}c@{}}Open rate rel. \\ to baseline\end{tabular}  & P-value  \\
\hline
(1) PPR  & 2.01 & $.00011^{***}$ (vs. baseline)\\
\hline
(2) CPR &  2.89 & $.00036^{***}$ (vs. 1) \\
\hline
(3) CPR with MM k=1 & 3.12 & $.00073^{***}$ (vs. 2) \\
\hline
(4) CPR with MM k= 4 & 3.22 & $.0046^{**}$ (vs. 3) \\
\hline
\end{tabular}
\caption{Open rates of different methods and statistical significance tests.}
\label{or}
\vspace{-0.5cm}
\end{table}

\section{Conclusion and Future Works}
This paper presents a mixture model for post-purchase complementary product recommendation. The model learns different open rate prediction weights for different user-product contexts. Experiment results show that CPR is superior than PPR by a significant margin, and the proposed mixture model gives additional gain in open rates. There are several possibilities to extend the research. Temporal features of user behavior are not included in the model. Furthermore, it will also be beneficial to consider product complementary score at different time scales, so that complementary product can be pushed to users at the most appropriate time.

\vspace{-0.0cm}
\bibliographystyle{ACM-Reference-Format}
\bibliography{refs} 


\begin{thebibliography}{00}

%%% the Web address.

\ifx \showCODEN    \undefined \def \showCODEN     #1{\unskip}     \fi
\ifx \showDOI      \undefined \def \showDOI       #1{{\tt DOI:}\penalty0{#1}\ }
  \fi
\ifx \showISBNx    \undefined \def \showISBNx     #1{\unskip}     \fi
\ifx \showISBNxiii \undefined \def \showISBNxiii  #1{\unskip}     \fi
\ifx \showISSN     \undefined \def \showISSN      #1{\unskip}     \fi
\ifx \showLCCN     \undefined \def \showLCCN      #1{\unskip}     \fi
\ifx \shownote     \undefined \def \shownote      #1{#1}          \fi
\ifx \showarticletitle \undefined \def \showarticletitle #1{#1}   \fi
\ifx \showURL      \undefined \def \showURL       #1{#1}          \fi
% invisible to TeX
\providecommand\bibfield[2]{#2}
\providecommand\bibinfo[2]{#2}
\providecommand\natexlab[1]{#1}
\providecommand\showeprint[2][]{arXiv:#2}

\bibitem[\protect\citeauthoryear{Bell, Bleau, and Davey}{Bell
  et~al\mbox{.}}{2011}]%
        {bell2011push}
\bibfield{author}{\bibinfo{person}{Kris~M Bell}, \bibinfo{person}{Darryl~N
  Bleau}, {and} \bibinfo{person}{Jeffrey~T Davey}.}
  \bibinfo{year}{2011}\natexlab{}.
\newblock \bibinfo{title}{Push notification service}.
\newblock   (\bibinfo{date}{Nov.~22} \bibinfo{year}{2011}).
\newblock
\newblock
\shownote{US Patent 8,064,896.}


\bibitem[\protect\citeauthoryear{Bilmes et~al\mbox{.}}{Bilmes
  et~al\mbox{.}}{1998}]%
        {bilmes1998gentle}
\bibfield{author}{\bibinfo{person}{Jeff~A Bilmes} {and}
  \bibinfo{person}{others}.} \bibinfo{year}{1998}\natexlab{}.
\newblock \showarticletitle{A gentle tutorial of the EM algorithm and its
  application to parameter estimation for Gaussian mixture and hidden Markov
  models}.
\newblock \bibinfo{journal}{{\em International Computer Science Institute\/}}
  \bibinfo{volume}{4}, \bibinfo{number}{510} (\bibinfo{year}{1998}),
  \bibinfo{pages}{126}.
\newblock


\bibitem[\protect\citeauthoryear{Bobadilla, Ortega, Hernando, and
  Guti{\'e}rrez}{Bobadilla et~al\mbox{.}}{2013}]%
        {bobadilla2013recommender}
\bibfield{author}{\bibinfo{person}{Jes{\'u}s Bobadilla},
  \bibinfo{person}{Fernando Ortega}, \bibinfo{person}{Antonio Hernando}, {and}
  \bibinfo{person}{Abraham Guti{\'e}rrez}.} \bibinfo{year}{2013}\natexlab{}.
\newblock \showarticletitle{Recommender systems survey}.
\newblock \bibinfo{journal}{{\em Knowledge-based systems\/}}
  \bibinfo{volume}{46} (\bibinfo{year}{2013}), \bibinfo{pages}{109--132}.
\newblock


\bibitem[\protect\citeauthoryear{Cetintas, Chen, and Si}{Cetintas
  et~al\mbox{.}}{2013a}]%
        {cetintas2013forecasting}
\bibfield{author}{\bibinfo{person}{Suleyman Cetintas}, \bibinfo{person}{Datong
  Chen}, {and} \bibinfo{person}{Luo Si}.} \bibinfo{year}{2013}\natexlab{a}.
\newblock \showarticletitle{Forecasting user visits for online display
  advertising}.
\newblock \bibinfo{journal}{{\em Information retrieval\/}}
  \bibinfo{volume}{16}, \bibinfo{number}{3} (\bibinfo{year}{2013}),
  \bibinfo{pages}{369--390}.
\newblock


\bibitem[\protect\citeauthoryear{Cetintas, Si, Xin, and Tzur}{Cetintas
  et~al\mbox{.}}{2013b}]%
        {cetintas2013probabilistic}
\bibfield{author}{\bibinfo{person}{Suleyman Cetintas}, \bibinfo{person}{Luo
  Si}, \bibinfo{person}{Yan~Ping Xin}, {and} \bibinfo{person}{Ron Tzur}.}
  \bibinfo{year}{2013}\natexlab{b}.
\newblock \showarticletitle{Probabilistic latent class models for predicting
  student performance}. \bibinfo{journal}{{\em CIKM, 2013\/}}
  (\bibinfo{year}{2013}), \bibinfo{pages}{1513--1516}.
\newblock


\bibitem[\protect\citeauthoryear{Herlocker, Konstan, Borchers, and
  Riedl}{Herlocker et~al\mbox{.}}{1999}]%
        {herlocker1999algorithmic}
\bibfield{author}{\bibinfo{person}{Jonathan~L Herlocker},
  \bibinfo{person}{Joseph~A Konstan}, \bibinfo{person}{Al Borchers}, {and}
  \bibinfo{person}{John Riedl}.} \bibinfo{year}{1999}\natexlab{}.
\newblock \showarticletitle{An algorithmic framework for performing
  collaborative filtering}. In \bibinfo{booktitle}{{\em SIGIR, 1999}}. ACM,
  \bibinfo{pages}{230--237}.
\newblock


\bibitem[\protect\citeauthoryear{Jin, Si, and Zhai}{Jin et~al\mbox{.}}{2006}]%
        {jin2006study}
\bibfield{author}{\bibinfo{person}{Rong Jin}, \bibinfo{person}{Luo Si}, {and}
  \bibinfo{person}{Chengxiang Zhai}.} \bibinfo{year}{2006}\natexlab{}.
\newblock \showarticletitle{A study of mixture models for collaborative
  filtering}.
\newblock \bibinfo{journal}{{\em Information Retrieval\/}} \bibinfo{volume}{9},
  \bibinfo{number}{3} (\bibinfo{year}{2006}), \bibinfo{pages}{357--382}.
\newblock


\bibitem[\protect\citeauthoryear{Kumar and Johari}{Kumar and Johari}{2015}]%
        {kumar2015push}
\bibfield{author}{\bibinfo{person}{Arvind Kumar} {and} \bibinfo{person}{Suchi
  Johari}.} \bibinfo{year}{2015}\natexlab{}.
\newblock \showarticletitle{Push notification as a business enhancement
  technique for e-commerce}. In \bibinfo{booktitle}{{\em ICIIP, 2015}}. IEEE,
  \bibinfo{pages}{450--454}.
\newblock


\bibitem[\protect\citeauthoryear{Lu, Chen, and Lakshmanan}{Lu
  et~al\mbox{.}}{2015}]%
        {lu2015competition}
\bibfield{author}{\bibinfo{person}{Wei Lu}, \bibinfo{person}{Wei Chen}, {and}
  \bibinfo{person}{Laks~VS Lakshmanan}.} \bibinfo{year}{2015}\natexlab{}.
\newblock \showarticletitle{From competition to complementarity: comparative
  influence diffusion and maximization}.
\newblock \bibinfo{journal}{{\em Proceedings of the VLDB Endowment\/}}
  \bibinfo{volume}{9}, \bibinfo{number}{2} (\bibinfo{year}{2015}),
  \bibinfo{pages}{60--71}.
\newblock


\bibitem[\protect\citeauthoryear{McAuley, Pandey, and Leskovec}{McAuley
  et~al\mbox{.}}{2015}]%
        {mcauley2015inferring}
\bibfield{author}{\bibinfo{person}{Julian McAuley}, \bibinfo{person}{Rahul
  Pandey}, {and} \bibinfo{person}{Jure Leskovec}.}
  \bibinfo{year}{2015}\natexlab{}.
\newblock \showarticletitle{Inferring networks of substitutable and
  complementary products}. In \bibinfo{booktitle}{{\em SIGKDD, 2015}}. ACM,
  \bibinfo{pages}{785--794}.
\newblock


\bibitem[\protect\citeauthoryear{Moertini and Nugroho}{Moertini and
  Nugroho}{2012}]%
        {moertini2012commerce}
\bibfield{author}{\bibinfo{person}{Veronica~S Moertini} {and}
  \bibinfo{person}{Criswanto~D Nugroho}.} \bibinfo{year}{2012}\natexlab{}.
\newblock \showarticletitle{E-commerce mobile marketing model resolving users
  acceptance criteria}.
\newblock \bibinfo{journal}{{\em International Journal of Managing Information
  Technology\/}} \bibinfo{volume}{4}, \bibinfo{number}{4}
  (\bibinfo{year}{2012}), \bibinfo{pages}{23}.
\newblock


\bibitem[\protect\citeauthoryear{Su and Khoshgoftaar}{Su and
  Khoshgoftaar}{2009}]%
        {su2009survey}
\bibfield{author}{\bibinfo{person}{Xiaoyuan Su} {and} \bibinfo{person}{Taghi~M
  Khoshgoftaar}.} \bibinfo{year}{2009}\natexlab{}.
\newblock \showarticletitle{A survey of collaborative filtering techniques}.
\newblock \bibinfo{journal}{{\em Advances in artificial intelligence\/}}
  \bibinfo{volume}{2009} (\bibinfo{year}{2009}), \bibinfo{pages}{4}.
\newblock


\bibitem[\protect\citeauthoryear{Tan, Roegiest, Lin, and Clarke}{Tan
  et~al\mbox{.}}{2016}]%
        {tan2016exploration}
\bibfield{author}{\bibinfo{person}{Luchen Tan}, \bibinfo{person}{Adam
  Roegiest}, \bibinfo{person}{Jimmy Lin}, {and} \bibinfo{person}{Charles~LA
  Clarke}.} \bibinfo{year}{2016}\natexlab{}.
\newblock \showarticletitle{An exploration of evaluation metrics for mobile
  push notifications}. In \bibinfo{booktitle}{{\em SIGIR, 2016}}. ACM,
  \bibinfo{pages}{741--744}.
\newblock


\bibitem[\protect\citeauthoryear{Tao and Zhai}{Tao and Zhai}{2004}]%
        {tao2004two}
\bibfield{author}{\bibinfo{person}{Tao Tao} {and} \bibinfo{person}{ChengXiang
  Zhai}.} \bibinfo{year}{2004}\natexlab{}.
\newblock \showarticletitle{A two-stage mixture model for pseudo feedback}. In
  \bibinfo{booktitle}{{\em SIGIR, 2004}}. ACM, \bibinfo{pages}{486--487}.
\newblock


\end{thebibliography}

\end{document}